\documentstyle[11pt,epsfig]{article}
\begin{document}
\title{\Large\bf Implication of BaBar's new data on the $D_{s1}(2710)$ and $D_{sJ}(2860)$}

\author{\small De-Min Li $^{1,2}$ and Bing Ma $^{1}$\\
\small   $^{1} $ Department of Physics, Zhengzhou University,
Zhengzhou, Henan 450052, People's Republic of China\\
\small $^{2}$ Theoretical Physics Center for Science Facilities,
CAS, Beijing 10049, Peoples's Republic of China }

\date{\today}

\maketitle

\vspace{0.5cm}

\begin{abstract}

The strong decays of the $D_{s1}(2710)$ and $D_{sJ}(2860)$ are
investigated in the framework of the $^3P_0$ model. Its decay
properties newly reported by the BaBar Collaboration can be
reasonably accounted for in the presence of the $D_{s1}(2710)$ being
a mixture of the $D_s(2\,^3S_1)$ and $D_s(1\,^3D_1)$. The orthogonal
partner of the $D_{s1}(2710)$ is expected to have a mass of about
$2.66\sim 2.9$ GeV in quark models and a width of about $40\sim 60$
MeV in the $^3P_0$ model. The predicted decay properties turn out to
be consistent with the BaBar's new data in both the orthogonal
partner of the $D_{s1}(2710)$ and the $D_s(1\,^3D_3)$
interpretations for the $D_{sJ}(2860)$. The available experimental
information is not enough to distinguish these two possibilities.
The $E1$ radiative transitions of the $D_{s1}(2710)$ and
$D_{sJ}(2860)$ are also studied. We tend to conclude that the
$D_{s1}(2710)$ can be identified as a mixture of the $D_s(2\,^3S_1)$
and $D_s(1\,^3D_1)$, and the $D_{sJ}(2860)$
 could be either the orthogonal partner of the
$D_{s1}(2710)$ or the $D_s(1\,^3D_3)$. Further experimental
information on the $D_{sJ}(2860)$ in the $D_s\eta$, $D^\ast_s\eta$,
and $DK^\ast$ channels is needed.

\end{abstract}

\vspace{0.5cm}

 {\bf PACS numbers:}14.40.Lb, 12.39.Jh, 13.25.Ft, 13.40.Hq

\newpage

\baselineskip 24pt

\section*{I. Introduction}
\indent \vspace*{-1cm}

In 2006, two new charm-strange states $D_{sJ}(2860)$ [Mass:
$2856.6\pm 1.5\pm 5.0$ Mev, Width: $48\pm7\pm 10$ MeV ] and
$D_{sJ}(2688)$ [Mass: $2688\pm 4\pm 3$ Mev, Width: $112\pm7\pm 36$
MeV ] were observed by the BaBar Collaboration in the $DK$
channel\cite{babar1}. There is no evidence for the $D_{sJ}(2688)$
and $D_{sJ}(2860)$ in the $D^\ast K$ or $D_s\eta$ mode and hence
their possible $J^P$ quantum numbers can be $0^+,~1^-,2^+,~3^-$,
etc. Subsequently, the Belle Collaboration observed a vector state
$D_{s1}(2710)$ [Mass: $2708\pm 9\,^{+11}_{-10} $ Mev, Width: $108\pm
23\,^{+36}_{-31}$ MeV ] in the $DK$ channel\cite{belle1}. Since
their reported masses and widths are consistent with each other, the
$D_{sJ}(2688)$ and $D_{s1}(2710)$ are believed to refer to a single
state with a mass of $2690\pm 7$ MeV and a width of $110\pm 27$
MeV\cite{pdg08}.

 More recently, the
$D_{sJ}(2860)$ and $D_{s1}(2710)$ were found again by BaBar
Collaboration in both $DK$ and $D^\ast K$ channels\cite{babar2}. The
resulting masses and widths of these two states are
\begin{eqnarray}
&&M(D_{sJ}(2860)^+)=2862\pm 2^{+5}_{-2}~\mbox{MeV},
\Gamma(D_{sJ}(2860)^+)=48\pm 3\pm 6 ~\mbox{MeV},\\
&&M(D_{s1}(2710)^+)=2710\pm 2^{+12}_{-7}~\mbox{MeV},
\Gamma(D_{s1}(2710)^+)=149\pm 7^{+39}_{-52}~ \mbox{MeV},
\end{eqnarray}
and the following ratios of branching fractions were also obtained :
\begin{eqnarray}
\frac{{\cal{B}}(D_{s1}(2710)^+\rightarrow D^\ast
K)}{{\cal{B}}(D_{s1}(2710)^+\rightarrow D K)}=0.91\pm
0.13\pm 0.12,\label{rate1}\\
\frac{{\cal{B}}(D_{sJ}(2860)^+\rightarrow D^\ast
K)}{{\cal{B}}(D_{sJ}(2860)^+\rightarrow D K)}=1.10\pm 0.15\pm
0.19.\label{rate2}
\end{eqnarray}
The observation of the $D_{sJ}(2860)$ in both $DK$ and $D^\ast K$
channels rules out it to be the $0^+$ state, since a $^3P_0$
$c\bar{s}$ state is forbidden to decay into $D^\ast K$. Its possible
spin-parity should be $1^-,~2^+,~3^-$, etc, if the structure at 2.86
GeV observed by the BaBar Collaboration in the $DK$ and $D^\ast K$
channels refer to a single resonance~\footnote{The two largely
overlapping resonances, namely a pair of radially excited tensor and
scalar $c\bar{s}$ states,  might also exist at about 2.86 GeV, as
proposed by Van Beveren and Rupp \cite{brupp}.}.

Apart from the $D_{s1}(2710)$ and $D_{sJ}(2860)$, in the $D^\ast K$
channel the BaBar Collaboration also found the evidence for the
$D_{sJ}(3040)$ whose mass and decay properties have been discussed
recently in Refs.\cite{zhangai,xiu,zhao1}. Here, we shall focus on
the implications of the BaBar's new data on the $D_{s1}(2710)$ and
$D_{sJ}(2860)$. Our main purpose is to check whether the
experimental data for these two states can be reasonably accounted
for in a simple $c\bar{s}$ picture or not. Therefore, the more
complex pictures such as multiquark configuration\cite{wang,vijande}
and two-state structure\cite{brupp,zhao1} are not adopted.

Systematic studies on the heavy-light meson spectra in quark models
show that the expected $D_s$ masses are about $2.66\sim 2.8$ GeV for
$1^-[D_s(2\,^3S_1)]$, $2.7\sim 2.9$ GeV for $1^-[D_s(1\,^3D_1)]$,
$2.8\sim 3.0$ for $3^-[D_s(1\,^3D_3)]$, $3.0\sim 3.2$ GeV for
$2^+[D_s(2\,^3P_2)]$, and $3.1\sim 3.3$ GeV for
$2^+[D_s(1\,^3F_2)]$,
respectively\cite{th4,th5,th6,th7,th8,th9,chang,mat,th10,close1}.
Only from mass, the $J^P=1^-$ assignment for the $D_{s1}(2710)$ is
strongly favored, consistent with the experiment. It therefore would
be most plausible that the $D_{sJ}(2860)$ is assigned to be either a
$1^-$ or a $3^-$ state. In the pure $S$-wave or $D$-wave $c\bar{s}$
picture, the decay properties of the $D_{s1}(2710)$ and
$D_{sJ}(2860)$ have been studied for several different quantum
numbers using various
approaches\cite{close1,rupp,col1,col2,zhang,wei,zhao}. Further
theoretical efforts are still required in order to explain all the
data for decays of the $D_{s1}(2710)$ and $D_{sJ}(2860)$
satisfactorily\cite{fazio}.

As mentioned above, the pure $D_s(2\,^3S_1)$ and $D_s(1\,^3D_1)$
have the same $J^P$ and similar masses, in general, they can mix to
produce two physical $1^-$ states lying in the mass region of about
$2.66\sim 2.9$ GeV. Therefore, the observed $D_{s1}(2710)$ is most
likely a mixture of the $D_s(2\,^3S_1)$ and
$D_s(1\,^3D_1)$\cite{close1,li}. If this picture is reasonable, a
natural question is whether the $D_{sJ}(2860)$ can be described as
the orthogonal partner of the $D_{s1}(2710)$ or not, since in the
mass region of about $2.66\sim 2.9$ GeV, only the $D_{sJ}(2860)$ is
a plausible $1^-$ charm-strange candidate at the present
time~\footnote{In 2004, the $D_{sJ}(2632)$ was reported by the SELEX
Collaboration in the final states $D^+_s\eta$ and
$D^0K^+$\cite{selex}. Various possible interpretations for the
$D_{sJ}(2632)$ turn out to be very unlikely\cite{ds1}. It was even
regarded to be an experimental artefact\cite{ds1,ds2}. Therefore, we
don't consider the possibility of the $D_{sJ}(2632)$ being the $1^-$
$c\bar{s}$\cite{chao1}, although its mass is close to the quark
model prediction of about 2.66 GeV for the
$D_s(2\,^3S_1)$\cite{chang}. }. The masses of the $D_s(2\,^3P_2)$
and $D_s(1\,^3F_2)$ are much higher than 2.86 GeV, we therefore will
only focus on the $1^-$ and $3^-$ assignments for the
$D_{sJ}(2860)$. In the present work we try to clarify (i) the
possibility of the $D_{s1}(2710)$ and the $D_{sJ}(2860)$ being in
fact the mixtures of the $D_s(2\,^3S_1)$ and $D_s(1\,^3D_1)$ and
(ii) whether the $D_{sJ}(2860)$ can be assigned to be a
$D_s(1\,^3D_3)$ or not by comparing the $^3P_0$ model predictions
for their strong decays with the available experimental data.

The organization of this paper is as follows. In Sec. II, we discuss
the strong decays of the $D_{s1}(2710)$ and $D_{sJ}(2860)$ for
different possible assignments. The radiative transitions of these
two states are given in Sec. III. The summary and conclusion are
given in Sec. IV.

\section*{II. Strong decays of the $D_{s1}(2710)$ and $D_{sJ}(2860)$ }

\indent \vspace*{-1cm}

 In the scenario of the $D_{s1}(2710)$
being the mixture of the pure $D_s(2\,^3S_1)$ and $D_s(1\,^3D_1)$,
the eigenvectors of $D_{s1}(2710)$ and its orthogonal partner
$D_{s1}(M_X)$ can be written as
\begin{eqnarray}
&&|D_{s1}(2710)\rangle=\cos\theta |2\,^3S_1\rangle-\sin\theta
|1\,^3D_1\rangle,\label{mix1}\\
&&|D_{s1}(M_X)\rangle=\sin\theta |2\,^3S_1\rangle+\cos\theta
|1\,^3D_1\rangle,\label{mix2}
\end{eqnarray}
where the $\theta$ is the mixing angle and $M_X$ denotes the mass of
the physical state $D_{s1}(M_X)$. We apply Eqs.(\ref{mix1}) and
(\ref{mix2}) to $c\bar{s}$ rather than $s\bar{c}$ states. When one
applies these equations to $s\bar{c}$ states, the mixing angle would
be opposite in sign to our $\theta$.

In this work, we shall employ the $^3P_0$ model to evaluate the
tow-body open-flavor strong decays of the initial state. The $^3P_0$
model, also known as the quark pair creation model, has been
extensively applied to evaluate the strong decays of mesons from
light $q\bar{q}$ to heavy $c\bar{b}$, since it gives a considerably
good description of many of the observed decay amplitudes and
partial widths of hadrons. Some detailed reviews on the $^3P_0$
model can be found in Refs.\cite{3p0rev1,3p0rev2,3p0rev3,3p0rev4}.
Also, the simple harmonic oscillator (SHO) approximation for the
meson space wave functions is used in the strong decays
computations. This is typical of strong decay calculations and it
has been demonstrated that using the more realistic wave functions,
such as those obtained from Coulomb, plus the linear potential
model, does not change the results
significantly\cite{sho1,sho2,sho3}.

In the $^3P_0$ model, the partial widths of the $D_{s1}(2710)$ and
$D_{s1}(M_X)$ can be given by
\begin{eqnarray}
&&\Gamma(D_{s1}(2710)\rightarrow BC)=\frac{\pi
P}{4M^2_{D_{s1}(2710)}}\sum_{LS}|\cos\theta
{\cal{M}}^{LS}_{D_s(2\,^3S_1)\rightarrow
BC}-\sin\theta{\cal{M}}^{LS}_{D_s(1\,^3D_1)\rightarrow BC}|^2,\label{decay1}\\
&&\Gamma(D_{s1}(M_X)\rightarrow BC)=\frac{\pi
P}{4M^2_{D_{s1}(M_X)}}\sum_{LS}|\sin\theta
{\cal{M}}^{LS}_{D_s(2\,^3S_1)\rightarrow
BC}+\cos\theta{\cal{M}}^{LS}_{D_s(1\,^3D_1)\rightarrow
BC}|^2,\label{decay2}
\end{eqnarray}
where $B$ and $C$ denote the final state mesons, $P$ is the final
state momentum, and ${\cal{M}}^{LS}$ is the partial amplitude.
According to the explicit expression for ${\cal{M}}^{LS}$ given by
our previous work\cite{li1,li2,li4}, the input parameters include
the light $q\bar{q}$ production strength $\gamma$, the SHO wave
function scalar parameters $\beta$, and the constituent quarks
masses. The meson effective $\beta$ values used in this work are
shown in Table 1. These effective SHO $\beta$ values were obtained
by equating the root mean square radius of the SHO wave function to
that obtained from the simple nonrelativistic potential model
proposed by Lakhina and Swanson\cite{swanson}~\footnote{In the
procedure, the Mathematica program\cite{mathp} is used. }. In this
potential model, the zeroth-order Hamiltonian is
\begin{eqnarray}
H_0=\frac{\bf
{P}^2}{M_r}-\frac{4}{3}\frac{\alpha_s}{r}+br+\frac{32\alpha_s\sigma^3e^{-\sigma^2r^2}}{3\sqrt{\pi}m_qm_{\bar{q}}}
{\bf S}_q\cdot{{\bf S}_{\bar{q}}}, \label{h0}
\end{eqnarray}
where $M_r= 2m_qm_{\bar{q}}/(m_q+m_{\bar{q}})$; $m_q$ and ${\bf
S}_{q}$ ($m_{\bar{q}}$ and ${\bf S}_{\bar{q}}$) are the mass and
spin of the constituent quark $q$ (antiquark ${\bar{q}}$),
respectively; The parameters were chosen to reproduce reasonably the
masses of the low lying charm-strange states $D_s$, $D^\ast_s$,
$D_{s1}(2459)$, $D_{s1}(2535)$, $D_{s0}(2317)$, and $D_{s2}(2573)$
and are $\alpha_s=0.53$, $b=0.135$ $\mbox{GeV}^2$, $\sigma=1.13$
GeV. The constituent quarks masses are taken to be  $m_u=m_d=0.33$
GeV, $m_s=0.55$ GeV, and $m_c=1.45$ GeV. We take $\gamma=6.25$ by
fitting to the decay $D_{s2}(2573)\rightarrow DK+D^\ast
K+D_s\eta$~\footnote{Model: $\Gamma=20$ MeV, $\Gamma(D^\ast
K)/\Gamma(DK)=0.11$; PDG\cite{pdg08}: $\Gamma=20\pm 5$ MeV,
$\Gamma(D^\ast K)/\Gamma(DK)<0.33$.}.  The meson masses used to
determine the phase space and final state momenta are $M_K= 496$
MeV, $M_{\eta}=548$ MeV, $M_{K^\ast}=894$ MeV, $M_{D}=1867$ MeV,
$M_{D_s}=1969$ MeV, $M_{D^\ast}=2009$ MeV,
 $M_{D^\ast_s}=2112$ MeV, $M_{D_{s2}(2573)}=2573$ MeV, $M_{D_{s1}(2710)}=2710$ MeV, and  $M_{D_{sJ}(2860)}=2862$
 MeV. The meson flavor functions follow the conventions of Ref.\cite{li1}, for example, $D^+_s=-c\bar{s}$, $D^0=c\bar{u}$, $K^+=-u\bar{s}$, and
  $\eta=(u\bar{u}+d\bar{d})/2-s\bar{s}/\sqrt{2}$.
{\small
\begin{table}[hbt]
\begin{center}
\vspace*{-0.5cm}
 \caption{\small The meson effective $\beta$ values in MeV.}
 \vspace*{0.2cm}
\begin{tabular}{cccccc}\hline\hline
  $n\,^{2S+1}L_J$  & $u\bar{u}$ &$u\bar{s}$ & $s\bar{s}$ &$c\bar{u}$&$c\bar{s}$\\\hline

  $1\,^1S_0$&618&465&476&410  &489\\
  $1\,^3S_1$&267&296&337&340  &408\\
  $2\,^3S_1$&226&247&276&276  &323\\
  $1\,^3P_J$&248&269&299&297  &346\\
  $1\,^3D_J$&232&251&278&278  &318\\
  \hline\hline
\end{tabular}
\end{center}
\end{table}
}

 {\small
\begin{table}[hbt]
\begin{center}
\vspace*{-0.5cm} \caption{\small Partial widths of the
$D_{s1}(2710)$ and $D_{sJ}(2860)$ as the $D_s(2\,^3S_1)$ and
$D_s(1\,^3D_1)$ mixtures in MeV ( $c\equiv\cos\theta$,
$s\equiv\sin\theta$).}
 \vspace*{0.2cm}
\begin{tabular}{ccc}\hline\hline
              & \multicolumn{1}{c}{$D_{s1}(2710)$}&\multicolumn{1}{c}{$D_{sJ}(2860)$}\\
 Mode         & $\Gamma_i$ &$\Gamma_i$\\\hline
$DK$          & $4.4c^2-39.1cs+86.8s^2$   &$63.3c^2-8.1cs+0.3s^2$   \\
$D_s\eta$     & $0.8c^2-6.2cs+12.7s^2$   &$20.2c^2+3.9cs+0.2s^2$    \\
$D^\ast K$    & $34.9c^2+72.1cs+37.2s^2$  &$37.7c^2-45.1cs+13.5s^2$ \\
$D^\ast_s\eta$& $1.4c^2+2.9cs+1.5s^2$        &$8.5c^2-12.9cs+4.9s^2$\\
$DK^\ast$     && $37.9c^2-108.1cs+77.1s^2$\\\hline
              &$\Gamma_{\mbox{total}}=41.4c^2+29.6cs+138.2s^2$
              &$\Gamma_{\mbox{total}}=167.5c^2-170.2cs+96.0s^2$\\\hline\hline
\end{tabular}
\end{center}
\end{table}
}

\begin{figure}[hbt]
\begin{center}
\epsfig{file=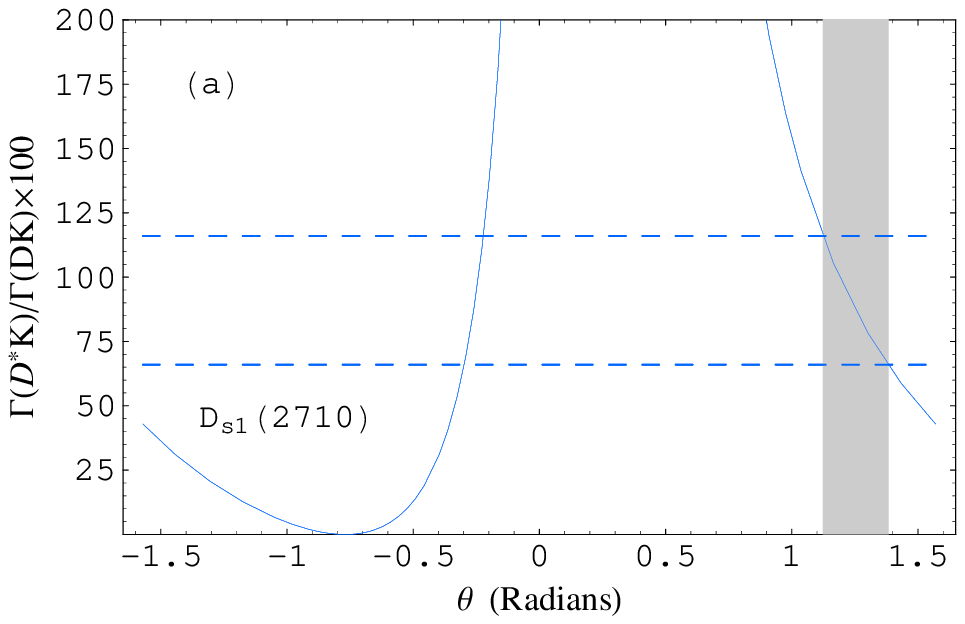,width=6.0cm, clip=}
\epsfig{file=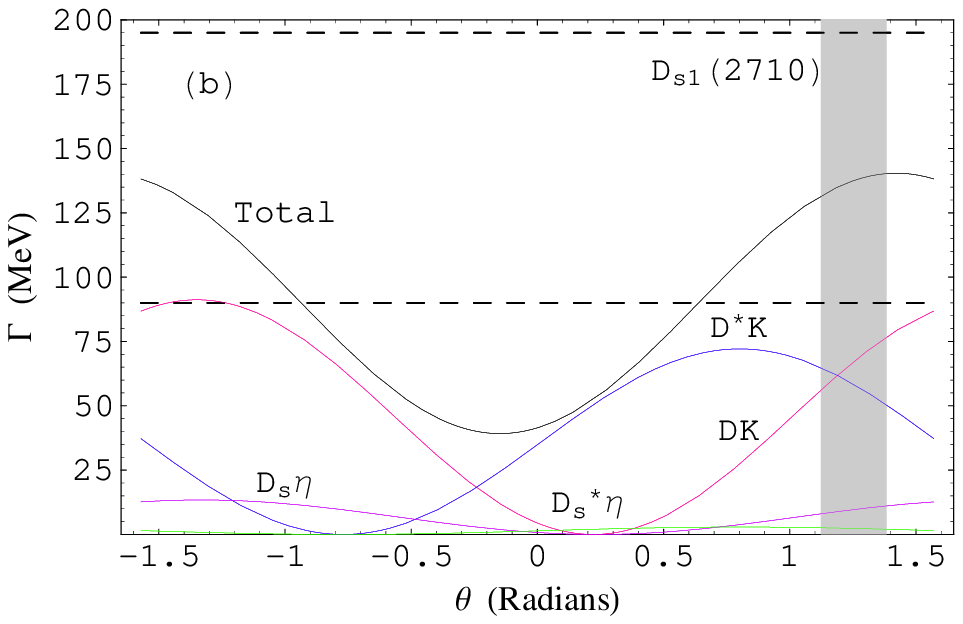,width=6.0cm,clip=}
\epsfig{file=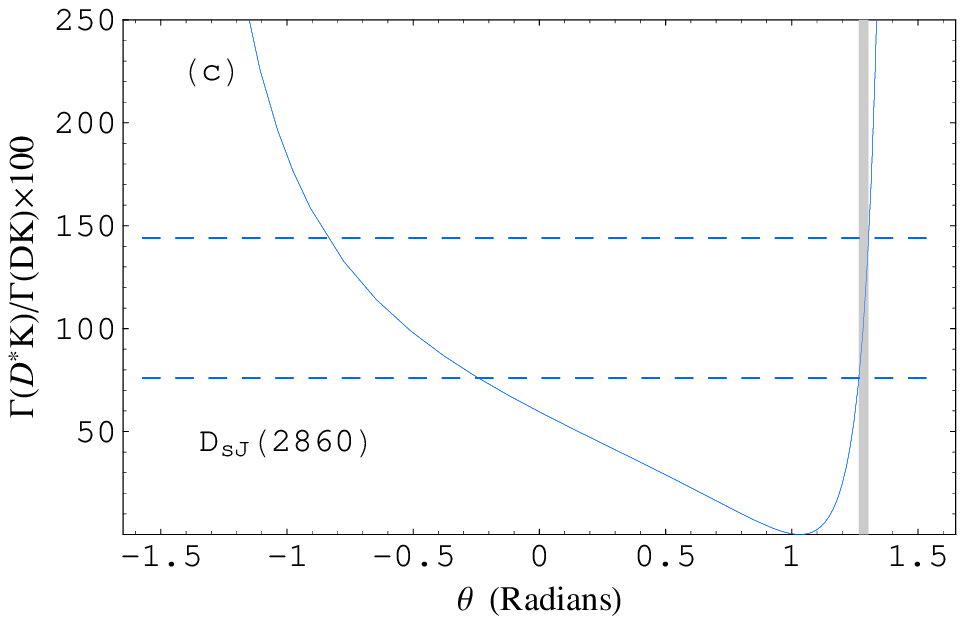,width=6.0cm, clip=}
\epsfig{file=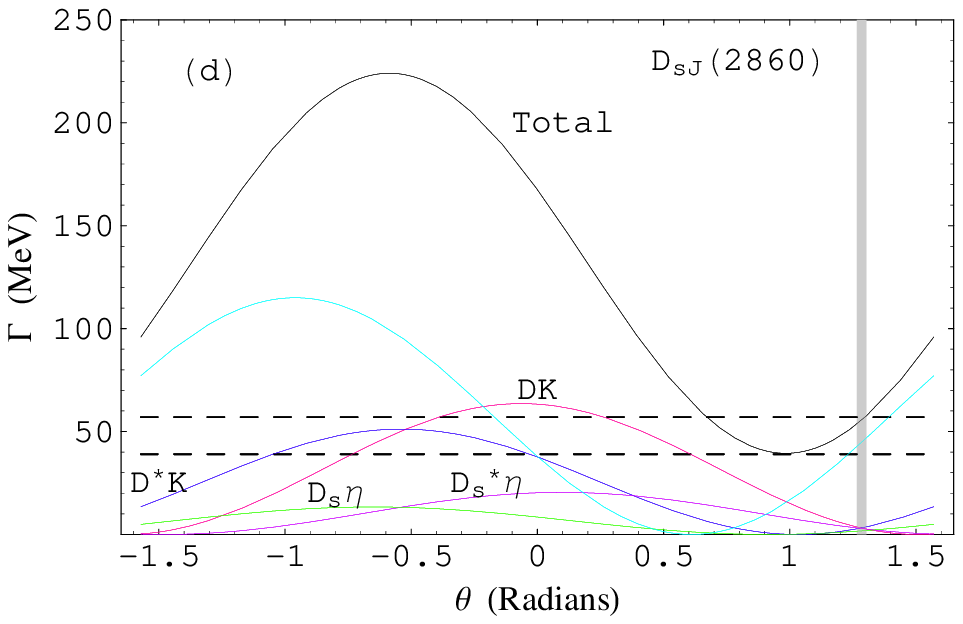,width=6.0cm, clip=}

 \vspace*{-0.3cm}
 \caption{\small Partial widths and $\Gamma(D^\ast
K)/\Gamma(DK)$ for the $D_{s1}(2710)$ and $D_{sJ}(2860)$ as
$J^P=1^-$ versus  $\theta$.  The horizontal dashed lines indicate
the upper and lower limits of the BaBar's data\cite{babar2}.}
\end{center}
\end{figure}

The numerical results for the partial widths of the $D_{s1}(2710)$
based on (\ref{decay1}) are listed in Table 2. The variation of
these partial widths and the ratio of $D^\ast K$ to $DK$ widths with
the mixing angle $\theta$ is shown in Fig. 1. From the experimental
result (\ref{rate1}), the mixing angle $\theta$ is found to
approximately satisfy (see Fig. 1 (a))
\begin{eqnarray}
1.12\leq\theta \leq 1.38 ~\mbox{radians}. \label{condition}
\end{eqnarray}
 It is obvious from Fig.
1(b) the total width of the $D_{s1}(2710)$ can be reasonably
reproduced in the presence of $1.12 \leq \theta \leq 1.38$ radians,
which therefore suggests that the picture of the $D_{s1}(2170)$
being in fact a mixture of the $D_s(2\,^3S_1)$ and $D_s(1\,^3D_1)$
appears reasonable. The quantitative calculations in a constituent
quark model with effective Lagrangians\cite{zhao1} also support this
picture.

In the presence of the $D_{s1}(2710)$ corresponding to one physical
state in the $D_s(2\,^3S_1)$-$D_s(1\,^3D_1)$ mixing scenario, from
condition (\ref{condition}) we can learn some decay information on
another physical state $D_{s1}(M_X)$.  Based on (\ref{decay2}), the
predicted total width and $\Gamma(D^\ast K)/\Gamma(DK)$ for the
$D_{s1}(M_X)$ are shown in Fig. 2 as functions of the initial state
mass $M_X$ and the mixing angle $\theta$. The $D_{s1}(M_X)$ mass is
restricted to be about $2.66\sim 2.9$ GeV expected by the quark
models. From Fig. 2, one can see that with the variations of the
initial state mass and the mixing angle, the total width of the
$D_{s1}(M_X)$ varies from about 40 to 60 MeV. The total width
depends weakly on $\theta$ while the ratio of $D^\ast K$ to $DK$
widths varies dramatically with $\theta$. The predicted width would
be helpful to search for and confirm another $1^-$ charm-strange
state in about $2.66\sim 2.9$ GeV experimentally.

\begin{figure}[hbt]
\begin{center}
\epsfig{file=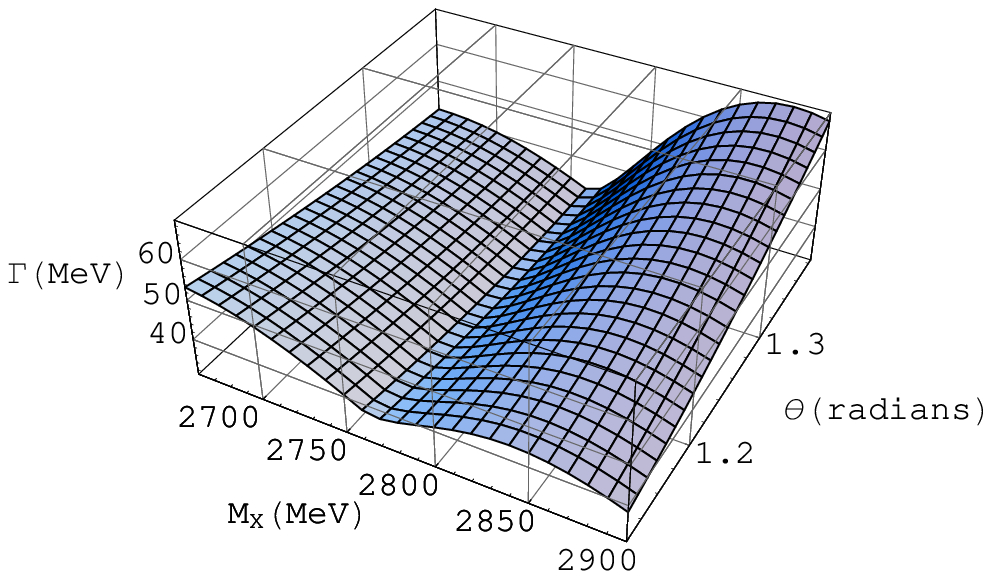,width=7.0cm, clip=}
\epsfig{file=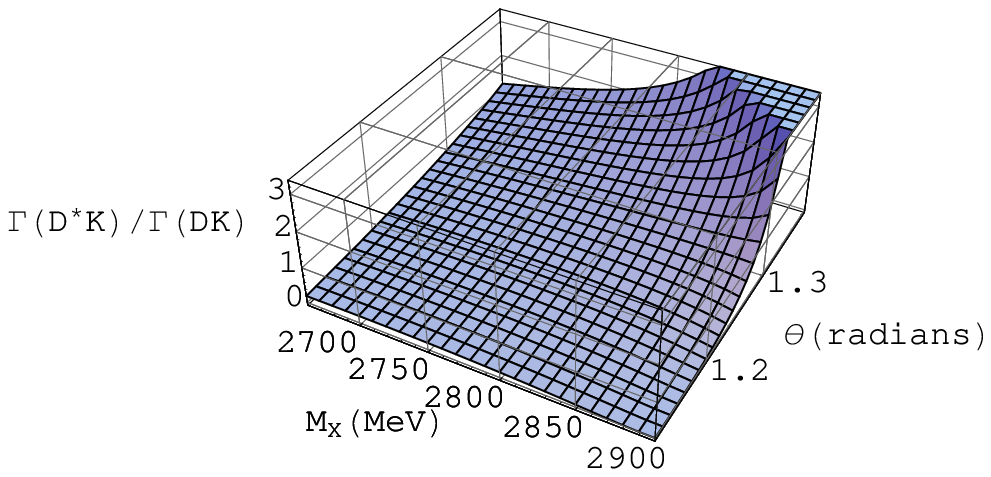,width=7.0cm,clip=}

 \vspace*{-0.4cm}
 \caption{\small Total width and $\Gamma(D^\ast
K)/\Gamma(DK)$ for the $D_{s1}(M_X)$ versus $M_X$ and the mixing
angle $\theta$. }
\end{center}
\end{figure}

We now turn to the possible assignments for the $D_{sJ}(2860)$. If
it is a $1^-$ state, as mentioned in Sec. I,  most likely it is the
orthogonal partner of the $D_{s1}(2710)$. Under this picture, the
numerical results for partial widths of the $D_{sJ}(2860)$ are
listed in Table 2 and the dependence of these partial widths as well
as $\Gamma(D^\ast K)/\Gamma(DK)$ on the mixing angle $\theta$ is
shown in Fig. 1. It is clear from Figs. 1(c) and 1(d) that both the
total width and $\Gamma(D^\ast K)/\Gamma(DK)$ for the $D_{sJ}(2860)$
can be properly reproduced with $1.26\leq \theta \leq 1.31$ radians,
just lying on the range of $1.12\leq \theta \leq 1.38$ radians.
Therefore, in the $^3P_0$ model, the possibility of the
$D_{sJ}(2860)$ being in fact the orthogonal partner of the
$D_{s1}(2710)$ does exist, if the $D_{sJ}(2860)$ is indeed a
$J^P=1^-$ $c\bar{s}$ state.

On the other hand, if the $D_{sJ}(2860)$ is a $3^-$ state, it would
be a natural candidate for the $D_s(1\,^3D_3)$ based on its mass. In
this case, the predicted partial widths are listed in Table 3. The
total width and the decay branching ratio fraction between $D^\ast
K$ and $DK$ modes are
\begin{eqnarray}
\Gamma\simeq 67~\mbox{MeV},~~ \frac{\Gamma(D^\ast
K)}{\Gamma(DK)}\simeq 0.8.
\end{eqnarray}
The predicted ratio of $D^\ast K$ to $DK$ widths is in agreement
with the experiment result (\ref{rate2}), and the total width is
also roughly consistent with the datum of $48\pm 7\pm 10$ MeV within
errors. Therefore, the $D_s(1\,^3D_3)$ interpretation for the
$D_{sJ}(2860)$ also seems acceptable. The recent lattice QCD study
also favors this assignment\cite{lattice}.
 {\small
\begin{table}[hbt]
\begin{center}
\vspace*{-0.5cm} \caption{\small Partial widths of the
$D_{sJ}(2860)$ as the $D_s(1\,^3D_3)$ in MeV. }
 \vspace*{0.2cm}
\begin{tabular}{ccccccc}\hline\hline
             $\Gamma(DK)$ & $\Gamma(D_s\eta)$ & $\Gamma(D^\ast K)$ & $\Gamma(D^\ast_s\eta)$& $\Gamma(DK^\ast)$&$\Gamma_{\mbox{total}}$  \\\hline\hline
             35.6& 1.6&26.8&0.6&2.7&67\\\hline\hline
\end{tabular}
\end{center}
\end{table}
}

The available experimental information on the $D_{sJ}(2860)$ is not
enough to distinguish the $D_s(1\,^3D_3)$ assignment from the
orthogonal partner of the $D_{s1}(2710)$ interpretation for the
$D_{sJ}(2860)$. However, the decay patterns for these two
assignments are very different. For example, for the
$D_s(1\,^3D_3)$, $\Gamma(D^\ast_s\eta)/\Gamma(DK)\simeq 0.02$,
$\Gamma(D_s\eta)/\Gamma(DK)\simeq 0.05$, and
$\Gamma(DK^\ast)/\Gamma(DK)\simeq 0.08$,  while for the orthogonal
partner of the $D_{s1}(2710)$,
$\Gamma(D^\ast_s\eta)/\Gamma(DK)\simeq 0.5$,
$\Gamma(D_s\eta)/\Gamma(DK)\simeq 0.9$, and
$\Gamma(DK^\ast)/\Gamma(DK)\simeq 13$. The further experimental
search of the $D_{sJ}(2860)$ in the  $D_s\eta$, $D^\ast_s\eta$, and
$DK^\ast$ channels would be crucial to distinguish the above two
possible assignments.

\section*{III. Radiative Transitions}
\indent\vspace*{-1cm}

 It is well known that radiative transitions can probe the
internal charge structure of hadrons, and therefore they will likely
play an important role in determining the quantum numbers and
hadronic structures of the $D_{s1}(2710)$ and $D_{sJ}(2860)$. In
this section, we shall evaluate the $E1$ transitions widths of the
$D_{s1}(2710)$ and $D_{sJ}(2860)$.

The partial width for the $E1$ transitions between the
$n~^{2S+1}L_J$ and $n^\prime\,^{2S^\prime+1}L^\prime_{J^\prime}$
$c\bar{s}$ states in the nonrelativistic quark model is given
by\cite{e1,closeq}
\begin{eqnarray}
\Gamma_{E1}(n\,^{2S+1}L_J\rightarrow
n^\prime\,^{2S^\prime+1}L^\prime_{J^\prime}+\gamma)=\frac{4}{3}\alpha
e^2_Q C_{fi}\delta_{SS^\prime}\left |\langle
n^\prime\,^{2S^\prime+1}L^\prime_{J^\prime}|r|n\,^{2S+1}L_J\rangle
\right |^2\frac{E^3_\gamma E_f}{M_i},
\end{eqnarray}
where $e_Q=\frac{2m_s-m_c}{3(m_s+m_c)}$, $\alpha=\frac{1}{137}$ is
the fine-structure constant, $E_\gamma$ is the final photon energy,
$E_f$ is the energy of the  final state
$n^\prime\,^{2S^\prime+1}L^\prime_{J^\prime}$, $M_i$ is the initial
state mass, and the angular matrix element $C_{fi}$ is
\begin{eqnarray}
C_{fi}=\mbox{Max}(L,L^\prime)(2J^\prime+1)\left\{\begin{array}{ccc}L^\prime&J^\prime&S\\
J&L&1\end{array}\right\}^2.
\label{cfi}
\end{eqnarray}

The wave functions used to evaluate the matrix element $\langle
n^\prime\,^{2S^\prime+1}L^\prime_{J^\prime}|r|n\,^{2S+1}L_J\rangle$
are obtained from the simple nonrelativistic quark model (\ref{h0}).
 Masses of the final states are
$M_{D_{s2}(2573)}=2573$ MeV, $M_{D_{s0}(2317)}=2317$ MeV,
$M_{D_{s1}(2460)}=2459$ MeV, and $M_{D{s1}(2536)}=2535$
MeV~\footnote{According to the PDG\cite{pdg08}, the well established
$c\bar{s}$ states include the $D_s(1968)$, $D^\ast_s(2112)$,
$D_{s0}(2317)$, $D_{s1}(2460)$, $D_{s1}(2536)$, and $D_{s2}(2573)$.
The $2\,^3S_1$, $1\,^3D_1$, and $1\,^3D_3$ $c\bar{s}$ are forbidden
to decay into $D_s(^3S_1)\gamma$ and $D_s(^1S_0)\gamma$ based on Eq.
(\ref{cfi}). Therefore, we only consider the processes where the
final states contain the $D_{s0}(2317)$, $D_{s1}(2460)$,
$D_{s1}(2536)$, and $D_{s2}(2573)$.}. The eigenvectors of the
physical states $D_{s1}(2460)$ and $D_{s1}(2536)$ are taken to be
\begin{eqnarray}
&&|D_{s1}(2459)\rangle=\cos\phi|^1P_1\rangle+\sin\phi|^3P_1\rangle,\\
&&|D_{s1}(2535)\rangle=-\sin\phi|^1P_1\rangle+\cos\phi|^3P_1\rangle,
\end{eqnarray}
where the mixing angle $\phi=-54.7^\circ$ as
Refs.\cite{swanson,closeq}. The resulting $E1$ transitions widths of
the the $D_{s1}(2710)$ and $D_{sJ}(2860)$ together with the photon
energies are given in Table 4. Without doubt, the experimental
studies on the processes such as $D_{sJ}(2860)\rightarrow
D_{s0}(2317)\gamma$, $D_{s1}(2459)\gamma$ and $D_{s1}(2535)\gamma$
will be helpful to distinguish the $D_s(1\,^3D_3)$ assignment from
the orthogonal partner of the $D_{s1}(2710)$ interpretation for the
$D_{sJ}(2860)$, since the decay modes of $D_s(^3P_0)\gamma$,
$D_s(^3P_1)\gamma$, and $D_s(^1P_1)\gamma$ are forbidden if the
$D_{sJ}(2860)$ is the $D_s(1\,^3D_3)$ while they are allowable if
the $D_{sJ}(2860)$ is the mixture of the $D_{s1}(2\,^3S_1)$ and
$D_{s1}(1\,^3D_1)$.

{\tiny
\begin{table}[hbt]
\begin{center}
\vspace*{-0.5cm} \caption{\small $E1$ transitions widths of the
$D_{s1}(2710)$ and $D_{sJ}(2860)$. ($E_\gamma$ in MeV, $\Gamma$ in
keV, $c\equiv\cos\theta$, and $s\equiv\sin\theta$). Estimates of
decay widths containing mixing angle $\theta$ are given in terms of
$1.26\leq \theta \leq 1.31$ radians extracted in Sec. II. A dash
indicates that a decay mode is forbidden based on Eq.(\ref{cfi}).}
 \vspace*{0.2cm}
\begin{tabular}{ccccccc}\hline\hline
      & \multicolumn{2}{c}{$D_{s1}(2710)$[S-D mixing]}&\multicolumn{2}{c}{$D_{sJ}(2860)$[S-D
      mixing]}&\multicolumn{2}{c}{$D_{sJ}(2860)$[$1^3D_3$]}\\
Final meson &$E_\gamma$& $\Gamma$& $E_\gamma$& $\Gamma$& $E_\gamma$&
$\Gamma$\\\hline

$D_{s2}(2573)$ &134 & $0.50c^2+0.19cs+0.02s^2=0.09\sim 0.12$&
274&$0.14c^2-1.53cs+4.13s^2=3.31\sim 3.48$&274 & 5.13\\

$D_{s0}(2317)$ &365 & $1.85c^2+6.88cs+6.39s^2=7.80\sim 7.97$&
493&$15.14c^2-16.29cs+4.38s^2=0.65\sim 1.04$&$-$ & $-$\\

$D_{s1}(2460)$ &239 & $1.10c^2+2.05cs+0.95s^2=1.47\sim 1.56$&375
&$3.48c^2-7.50cs+4.03s^2=1.80\sim 2.13$&$-$ & $-$\\

$D{s1}(2536)$ &169 & $0.20c^2+0.37cs+0.17s^2=0.27\sim 0.29$&
308&$0.99c^2-2.15cs+1.16s^2=0.52\sim 0.61$&$-$ & $-$\\\hline\hline
      \end{tabular}
\end{center}
\end{table}
}
\section*{IV. Summary and conclusion}
\indent \vspace*{-1cm}

The $D_{s1}(2710)$ has the definite $J^P=1^-$ and its mass is
similar to the quark models expectations for the masses of the pure
$2\,^3S_1$ and $1\,^3D_1$ $c\bar{s}$. Therefore, the $D_{s1}(2710)$
is most likely a mixture of the pure $2\,^3S_1$ and $1\,^3D_1$
$c\bar{s}$ states. We first check this possibility by studying its
strong decay properties in the $^3P_0$ model. Our calculations do
support this possibility. We also find the $1\,^3D_1$ component of
the $D_{s1}(2710)$ is large, in agreement with the recent lattice
QCD study\cite{lattice}. The orthogonal partner of the
$D_{s1}(2710)$ is predicted to have a width of about $40\sim 60$
MeV. This small decay width maybe result from the node structure in
its wave function where the large $2\,^3S_1$ component is expected
to exist. Also, the predicted width of about $40\sim 60$ MeV would
be helpful to search for and confirm another $1^-$ charm-strange
state in the mass region of about $2.66\sim 2.9$ GeV experimentally.

The observation of the $D_{sJ}(2860)$ in the $D^\ast K$ and $DK$
channels makes that it should have $J^P=1^-$, $2^+$, $3^-$,$\cdots$.
The constituent quark models predictions for the spectra of higher
$c\bar{s}$ states strongly favor that the $D_{sJ}(2860)$ is either a
$1^-$ or a $3^-$ state.

If $D_{sJ}(2860)$ is a $1^-$ $c\bar{s}$, its most plausible
assignment would be the orthogonal partner of the $D_{s1}(2710)$
based on its mass. In this picture, we evaluate the strong decay
pattern of the $D_{sJ}(2860)$. Our results indicate that all the
available data on its decays can be well explained. If the
$D_{sJ}(2860)$ is a $3^-$ $c\bar{s}$, from its mass, it would be a
good $1\,^3D_3$ candidate. The predicted decay pattern turn out to
be also consistent with the data. Therefore, both of these
assignments appear likely in the $^3P_0$ model.

The decay patterns of the above two assignments for the
$D_{sJ}(2860)$ are very different. The information on the partial
widths of $\Gamma(D_s\eta)$, $\Gamma(D^\ast_s\eta)$, and
$\Gamma(DK^\ast)$ is crucial to distinguish these two possibilities.
If further measurements refute these two possible assignments for
the $D_{sJ}(2860)$, the more complex interpretations such as the
tetraquark state or two-state structure would be really necessary.
Therefore, the further experimental search of the $D_{sJ}(2860)$ in
the $D_s\eta$, $D^\ast_s\eta$, and $DK^\ast$ channels is strongly
called for.

We don't consider the possibility of the $D_{sJ}(2860)$ being a
$2\,^3P_2$ or $1\,^3F_2$ $c\bar{s}$, since its mass is much lower
than the quark models predictions for the $2P$ and $1F$
charm-strange mesons. The results from a constituent quark with
effective Lagrangians\cite{zhao1} don't yet favor this possibility.

 We also investigate the $E1$ radiative transitions of the
$D_{s1}(2710)$ and $D_{sJ}(2860)$. The experimental studies on the
$E1$ transitions between $D_{sJ}(2860)$ and $1P$ charm-strange
mesons will also be helpful in determining the quantum numbers of
the $D_{sJ}(2860)$.

Comparing the predicted masses from the quark models and strong
decay properties from the $^3P_0$ model with the BaBar's new data on
the $D_{s1}(2710)$ and $D_{sJ}(2860)$, we tend to conclude that the
$D_{s1}(2710)$ can be identified as a mixture of the $D_s(2\,^3S_1)$
and $D_s(1\,^3D_1)$, and the $D_{sJ}(2860)$
 could be either the orthogonal partner of the
$D_{s1}(2710)$ or the $D_s(1\,^3D_3)$.

 \section*{Acknowledgments}
 \indent\vspace{-1cm}

 We acknowledge Prof. Bing-Song Zou for very helpful and stimulating discussions. This work
is supported in part by HANCET under Contract No. 2006HANCET-02,
and by the Program for Youthful Teachers in University of Henan
Province.


\baselineskip 18pt

\begin{thebibliography}{99}

\bibitem{babar1} B. Aubert et al.(BaBar Collaboration),
Phys. Rev. Lett. {\bf 97}, 222001 (2006) [arXiv:hep-ex/0607082].

\bibitem{belle1} K. Abe et al. (Belle Collaboration), arXiv:hep-ex/0608031;\\
J. Brodzicka et al. (Belle Collaboration), Phys. Rev. Lett.{\bf
100},092001 (2008) [arXiv:0707.3491 [hep-ex]].

\bibitem{pdg08} C. Amsler et al.(Partical Data Group), Phys. Lett. B {\bf 667}, 1
(2008).

\bibitem{babar2} B. Aubert et al.(Babar Collaboration), arXiv:0908.0806 [hep-ex].
\bibitem{brupp} E. Van Beveren and G. Rupp, arXiv:0908.1142 [hep-ph].

\bibitem{zhangai}B. Chen, D. X. Wang, and A. Zhang, Phys. Rev. D  {\bf
80}, 071502(R) (2009) [arXiv:0908.3261 [hep-ph]].

\bibitem{xiu} Z. F. Sun and X. Liu, Phys. Rev. D  {\bf 80}, 074037
(2009) [arXiv: 0909.1658 [hep-ph]].

\bibitem{zhao1} X. H. Zhong and Q. Zhao, arXiv:0911.1856 [hep-ph].


\bibitem{wang} Z. G. Wang, Chin. Phys. C {\bf 32}, 797 (2008) [arXiv:0708.0155
[hep-ph]].

\bibitem{vijande} J. Vijande, A. Valcarce, and F. Fernandez, Phys.
Rev. D {\bf 79}, 037501 (2009) [arXiv:0810.4988 [hep-ph]].

\bibitem{th4} S. Godfrey and N. Isgur, Phys. Rev. D {\bf 32}, 189
(1985).

\bibitem{th5} S. N. Gupta and J. M. Johnson, Phys. Rev. D {\bf 51},
168 (1995) [arXiv:hep-ph/9409432].

\bibitem{th6} J. Zeng, J. W. Van Orden, and W. Roberts, Phys. Rev. D
{\bf 52}, 5229 (1995) [arXiv:hep-ph/9412269].

\bibitem{th7} D. Ebert, V. O. Galkin and R. N. Faustov, Phys. Rev. D
{\bf 57}, 5663 (1998) [arXiv:hep-ph/9712318].

\bibitem{th8} T. A. Lahde, C. J. Nyfalt, and D. O. Riska, Nucl. Phys.
A {\bf 674}, 141 (2000) [arXiv:hep-ph/9908485].

\bibitem{th9} M. Di Pierro and E. Eichten, Phys. Rev. D {\bf 64},
114004 (2001) [arXiv:hep-ph/0104208].


\bibitem{chang} C. H. Chang, C. S. Kim, and G. L. Wang, Phys. Lett. B
{\bf 623}, 218 (2005) [arXiv:hep-ph/0505205].

\bibitem{mat} T. Matsuki, T. Morii, and K. Sudoh, Eur. Phys. J.
A {\bf 31}, 701 (2007) [arXiv:hep-ph/0610186].

\bibitem{th10} D. Ebert, R. N. Faustov, and V. O. Galkin,
arXiv:0910.5612 [hep-ph].

\bibitem{close1} F. E. Close, C. E. Thomas, O.
Lakhina, and E. S. Swanson, Phys. Lett. B {\bf 647},159 (2006)
 [arXiv:hep-ph/0608139].

\bibitem{rupp} E. van Beveren and G. Rupp, Phys. Rev. Lett. {\bf
97}, 202001 (2006) [arXiv:hep-ph/0606110].

\bibitem{col1} P. Colangelo, F. D Fazio, and S. Nicotri, Phys. Lett.
B {\bf 642}, 48 (2006) [arXiv:hep-ph/0607245].

\bibitem{col2} P. Colangelo, F. De Fazio, S. Nicotri, and M. Rizzi,
Phys. Rev. D {\bf 77}, 014012 (2008) [arXiv:0710.3068 [hep-ph]].


\bibitem{zhang} B. Zhang, X. Liu, W.Z. Deng, and S.L. Shu,
Eur. Phys. J. C {\bf 50}, 617 (2007) [arXiv:hep-ph/0609013].

\bibitem{wei} W. Wei, Xiang Liu, and Shi-Lin Zhu, Phys. Rev. D {\bf
75}, 014013 (2007) [arXiv:hep-ph/0612066].

\bibitem{zhao} X. H. Zhong and Q. Zhao, Phys. Rev. D {\bf 78},
014029 (2008) [arXiv:0803.2102 [hep-ph]].

\bibitem{fazio} F. De Fazio, arXiv:0910.0412 [hep-ph].

\bibitem{li} D. M. Li, B. Ma, and Y. H. Liu, Eur. Phys. J. C  {\bf
51}, 359 (2007) [arXiv:hep-ph/0703278].

\bibitem{selex} A. V. Evdokimov et al.(SELEX Collaboration), Phys.
Rev. Lett. {\bf 93}, 242001 (2004) [arXiv: hep-ex/0406045].

\bibitem{ds1} T. Barnes, F. E. Close, J. J. Dudek, S. Godfrey, and
E. S. Swanson, Phys. Lett. B {\bf 600}, 223 (2004) [arXiv:
hep-ph/0407120].

\bibitem{ds2} E. S. Swanson, Phys. Rept. {\bf 429}, 243  (2006) [arXiv:hep-ph/0601110].

\bibitem{chao1} K. T. Chao, Phys. Lett. B {\bf 599}, 43  (2004) [arXiv:hep-ph/0407091].


\bibitem{3p0rev1} A. Le Yaouanc, L. Oliver, O. Pene, and J-C. Raynal,
Hadron transitons in the quark model ( Gordon and Breach Science
Publishers, New York, 1988).

\bibitem{3p0rev2} W. Roberts and B. Silvestr-Brac, Few-Body Syst.
{\bf 11}, 171 (1992).

\bibitem{3p0rev3} H. G. Blundel, arXiv:hep-ph/9608473.

\bibitem{3p0rev4} E. S. Ackleh, T. Barnes, and E. S. Swanson,
Phys. Rev. D {\bf 54}, 6811 (1996) [arXiv:hep-ph/9604355];\\
T. Barnes, F. E. Close, P. R. Page, and E. S. Swanson, Phys. Rev. D
{\bf 55}, 4157 (1997) [arXiv:hep-ph/9609339];\\
 T. Barnes, N. Black, and P. R. Page, Phys. Rev. D
{\bf 68}, 054014 (2003) [arXiv:nucl-th/0208072].

\bibitem{sho1} R. Kokoski and N. Isgur, Phys. Rev. D  {\bf 35}, 907
(1987).

\bibitem{sho2} P. Geiger and E. S. Swanson, Phys. Rev. D {\bf 50},
6855 (1994) [arXiv:hep-ph/9405238].

\bibitem{sho3}H.G. Blundell and S. Godfrey, Phys. Rev. D {\bf
53},3700 (1996) [arXiv:hep-ph/9508264].

\bibitem{li1} D. M. Li and B. Ma, Phys. Rev. D {\bf 77}, 074004
(2008)[arXiv:0801.4821 [hep-ph]].

\bibitem{li2} D. M. Li and B. Ma, Phys. Rev. D {\bf 77}, 094021
(2008)[arXiv:0803.0106 [hep-ph]].


\bibitem{li4} D. M. Li and E. Wang, Eur. Phys. J. C  {\bf 63}, 297
(2009)[arXiv:0904.1252 [hep-ph]].


\bibitem{swanson} O. Lakhina and E. S. Swanson, Phys. Lett. B {\bf
650}, 159 (2007) [arXiv:hep-ph/0608011].

\bibitem{mathp} W. Lucha and F. F. Schoberl, Int. J. Mod. Phys. C
{\bf 10}, 607 (1999) [arXiv:hep-ph/9811453].


\bibitem{lattice} J. Koponen, Phys. Rev. D {\bf 78}, 074509  (2008) [arXiv:0708.2807 [hep-ph]].

\bibitem{e1} W. Kwong and J. L. Rosner, Phys. Rev. D {\bf 38}, 279
(1988);\\
S. Godfrey, Phys. Rev. D {\bf 70}, 054017 (2004) [arXiv:hep-ph/0406228];\\
S. Godfrey, Phys. Rev. D {\bf 72}, 054029 (2005)
 [arXiv:hep-ph/0508078].

\bibitem{closeq} F. E. Close and E. S. Swanson, Phys. Rev. D {\bf 72}, 094004
(2005) [arXiv:hep-ph/0505206].



\end{thebibliography}
\end{document}